\documentclass[11pt,showpacs,preprintnumbers,amsmath,amssymb]{revtex4}
\usepackage{amsfonts}
\usepackage{latexsym}
\usepackage{pxfonts}

\begin{document}

\title{Plasmon spectrum of degenerated electron gas. $T=0$ Green function
method. Detailed pedagogical derivation}
\author{Todor M. Mishonov}
\author{Liliya A. Atanasova}
\author{Peter A. Ivanov}
\author{Tihomir I. Valchev}
\author{Dimo L. Arnaudov}

\affiliation{Department of Theoretical Physics, Faculty of
Physics, Sofia University St.\,Kliment Ohridski, Bulgaria.}


\begin{abstract}
Plasmon spectrum and polarization operator of 1, 2, and 3
dimensional electron gas are calculated by $T=0$ Green function
technique. It is shown that this field theory method gives probably
the simplest pedagogical derivation of the statistical problem for
the response function. The explanation is complimentary to the
standard courses on condensed matter and plasma physics of the level
of IX volume of Landau-Lifshitz encyclopedia on theoretical physics.
\end{abstract}

\maketitle

\section{Introduction}
Plasma waves are well investigated in condensed matter physics. In
the present paper we derive the dispersion of plasma waves in the
framework of Green functions formalism for $T=0$. Our purpose is
purely pedagogical. We use the diagram technique described in a
textbook by Lifshitz and Pitaevskii \cite{Lifshitz9} in order to
derive the dielectric polarizability of degenerated electron gas.
The same result is derived in \cite{Lifshitz10} in a different way.
Our analysis starts with the $1D$ case. Then we consider the $2D$
case which is realized in semiconductor inverse layers. For a review
of the properties of two dimensional electron gas see, for example,
Ref.~\onlinecite{2DEG}.

\section{Model}
The dielectric formalism gives us the equation for the dielectric
polarizability
\begin{equation}
\varepsilon_l(\Omega,\mathbf q)=1-V(\Omega,\mathbf q)\Pi(\Omega,\mathbf q).
\end{equation}
In that expression $V(\Omega,\mathbf q)$ means photon propagator, which in
non-relativistic approximation
$$v_{\mathrm{gr}}=\frac{\partial\omega}{\partial k}\ll c$$
is the Fourier transformation of the Coulomb potential
\begin{equation}
V^{(2D)}(\Omega,\mathbf q) \approx \frac{2\pi e^2}{q}, \qquad
V^{(3D)}(\Omega,\mathbf q) \approx \frac{4\pi e^2}{q^2}.
\end{equation}

The polarization operator $\Pi$ is the functional derivative of the
electron density with respect to small perturbations of the electric
potential. The diagram rules described in Ref.~\cite{Lifshitz9}, section 13 give
for the simple electron loop
\begin{equation}
\Pi(\Omega,\mathbf q)=(-i)(-1)\int_P iG_{\alpha\beta}(P)iG_{\beta\alpha}(P+Q)
\frac{d^{D+1}P}{(2\pi)^{D+1}},
\end{equation}
where $G_{\alpha\beta}(P)$ are the Green functions of degenerated electron gas
\begin{equation}
G_{\alpha\beta}(P)=\delta_{\alpha\beta}G(P)=
\frac{\delta_{\alpha\beta}}{\omega-\eta_p(1-i0)}.
\end{equation}
We follow standard notations: $\eta_p=\varepsilon_p-\mu, \ \mu$ is the Fermi
energy, $\varepsilon_p=p^2/2m$ is the electron kinetic energy
and $P=(\omega,\mathbf p),\quad Q=(\Omega,\mathbf q)$.
The dispersion of longitudinal plasma waves $\Omega(q)$ is the solution of
\begin{equation}\label{eps=0}
\varepsilon_l(\Omega,\mathbf q)=0.
\end{equation}

The first step is calculating the polarization operator. Using
\begin{align}
G(P)G(P+Q)&=\frac{1}{\omega-\eta_p(1-i0)}.
\frac{1}{\omega+\Omega-\eta_{p+q}(1-i0)}=\\
&=\frac{1}{\eta_p-\eta_{p+q}+\Omega}
\left[\frac{1}{\omega-\eta_p(1-i0)}-
\frac{1}{\omega+\Omega-\eta_{p+q}(1-i0)}\right],\nonumber
\end{align}
the $\omega$ integration gives the electron filling factors
\begin{equation}
n_p=\frac{1}{2\pi i}\int\frac{d\omega}{\omega-\eta_p(1-i0)}=
\thetaup(p_F-p).
\end{equation}
In such a way we obtain
\begin{align}
\Pi&=-2i\int\frac{d^D p}{(2\pi)^D}\int\frac{d\omega}{2\pi}
G(P)G(P+Q)=\nonumber\\
&=2\int\frac{d^D p}{(2\pi)^D} \label{Pigen}
\frac{n_p-n_{p+q}}{\eta_p-\eta_{p+q}+\Omega + i0}=I_1+I_2\,,
\end{align}
where the evanescent imaginary part of the frequency corresponds to
the Landau rule for adiabatic appearance of the perturbation, see
Ref.~\cite{Lifshitz10} secs.~29, 40, Eqs.~(29.6), (40.15).
Here for brevity we use the notations
\begin{equation}
I_1(\Omega,\mathbf{q})=2\int\frac{d^D p}{(2\pi)^D}
\frac{n_p}{\eta_p-\eta_{p+q}+\Omega}=
2\int\frac{d^D p}{(2\pi)^D}
\frac{n_p}{\frac{p^2}{2m}-\frac{(\mathbf{p}+\mathbf{q})^2}{2m}+\Omega},
\end{equation}
\begin{align}
I_2(\Omega,\mathbf{q})&=2\int\frac{d^D p}{(2\pi)^D}
\frac{-n_{p+q}}{\eta_p-\eta_{p+q}+\Omega}=\nonumber
2\int\frac{d^D p}{(2\pi)^D}\frac{-n_{p+q}}
{\frac{p^2}{2m}-\frac{(\mathbf{p}+\mathbf{q})^2}{2m}+\Omega}=\\
&=2\int\frac{d^D p}{(2\pi)^D}
\frac{n_{p+q}}{\frac{(\mathbf{p}+\mathbf{q})^2}{2m}-
\frac{p^2}{2m}-\Omega}.
\end{align}
We change the variables $p'=p+q$. Then $dp=dp', \ n_{p+q}=n_{p'}$ and
\begin{equation}
I_2(\Omega,\mathbf{q})=2\int\frac{d^D p}{(2\pi)^D}
\frac{n_p}{\frac{p^2}{2m}-\frac{(\mathbf{p}-\mathbf{q})^2}{2m}-\Omega}=
I_1(-\Omega,-\mathbf{q}).
\end{equation}

The general formula \eqref{Pigen} for the response function of the
electron gas is given in many textbooks on statistical and solid
state physics.

In the next section we will analyze the simple case of $D=1$ which
is discussed in \cite{Stratton}.

\section{1-DIMENSIONAL CASE}
For the case of $D=1$ we assume $q\geq0$, and
\begin{align}\nonumber
I_1&=2\int_{-\infty}^{\infty}\frac{d p}{2\pi}
\frac{n_p}{\frac{p^2}{2m}-\frac{(p+q)^2}{2m}+\Omega}
=\frac{2m}{\pi}\int_{-\infty}^{\infty}
\frac{\thetaup(p_F-|p|)dp}{2m\Omega-q^2-2pq}=\\
&=\frac{2m}{\pi}\int_{-p_F}^{p_F}\frac{dp}{2m\Omega-q^2-2pq}
=-\frac{m}{\pi q}\int_{-p_F}^{p_F}\frac{dp}
{\left[p+\left(\frac{q}{2}-\frac{m\Omega}{q}\right)\right]-i0}.
\end{align}
Then
\begin{equation}
I_1=\frac{m}{\pi q}\ln\left|\frac{2m\Omega-q^2+2p_Fq}{2m\Omega-q^2-2p_Fq}
\right|-\frac{im}{q}\thetaup\left(1-\left|\frac{q}{2p_F}-\frac{m\Omega}{p_Fq}
\right|\right).
\end{equation}
The velocity on Fermi level is $v_F=p_F/m$, and
\begin{align}
I_1&=\frac{m}{\pi q}\ln\left|
\frac{\Omega-\frac{q^2}{2m}+v_Fq}{\Omega-\frac{q^2}{2m}-v_Fq}\right|
-\frac{im}{q}\thetaup\left(1-\left|\frac{q}{2p_F}-\frac{m\Omega}{p_Fq}
\right|\right),\\
I_2&=\frac{m}{\pi q}\ln\left|
\frac{\Omega+\frac{q^2}{2m}-v_Fq}{\Omega+\frac{q^2}{2m}+v_Fq}\right|
+\frac{im}{q}\thetaup\left(1-\left|\frac{q}{2p_F}
+\frac{m\Omega}{p_Fq}\right|\right).
\end{align}
Owing to the symmetry of the polarization operator
\begin{equation}\nonumber
\Pi(Q)=\Pi(-Q),
\end{equation}
we obtain
\begin{align}\nonumber
\Pi&=\frac{m}{\pi q}\ln\left|
\frac{(\Omega-\frac{q^2}{2m}+v_Fq)(\Omega
+\frac{q^2}{2m}-v_Fq)}{(\Omega-\frac{q^2}{2m}-v_Fq)
(\Omega+\frac{q^2}{2m}+v_Fq)}\right|-\frac{im}{q}\left[
\thetaup\left(1-\left|\frac{q}{2p_F}-\frac{m\Omega}{p_Fq}\right|\right)
-\thetaup\left(1-\left|\frac{q}{2p_F}+\frac{m\Omega}{p_Fq}\right|
\right)\right]=\\
&=\frac{m}{\pi q}\ln\left|
\frac{1-\left(\frac{\frac{q^2}{2m}-v_Fq}
{\Omega}\right)^2}{1-\left(\frac{\frac{q^2}{2m}+v_Fq}{\Omega}\right)^2}
\right|-\frac{im}{q}\left[\thetaup\left(1-\left|
\frac{q}{2p_F}-\frac{m\Omega}{p_Fq}\right|\right)
-\thetaup\left(1-\left|\frac{q}{2p_F}
+\frac{m\Omega}{p_Fq}\right|\right)\right].
\end{align}
In hydrodynamic approximation (long waves and phase velocity much
larger than Fermi velocity)
\begin{equation}\label{hydapp}
\frac{q^2}{2m}\ll\Omega\,,\ q\rightarrow0\,,\ v_F\ll\frac{\Omega}{q}=v_{\varphi}\,,
\end{equation}
we use
\begin{equation}\nonumber
(1+\epsilon)^n\approx1+n\epsilon\quad\textrm{for}\quad\epsilon\ll1.
\end{equation}
Then
\begin{align}\nonumber
\Pi&\approx\frac{m}{\pi q}\ln
\left|\left(1-\left(\frac{\frac{q^2}{2m}-v_Fq}{\Omega}\right)^2\right)
\left(1+\left(\frac{\frac{q^2}{2m}+v_Fq}{\Omega}\right)^2\right)\right|
\approx\\
&\approx\frac{m}{\pi q}\ln\left|1-\left(\frac{\frac{q^2}{2m}-v_Fq}
{\Omega}\right)^2+\left(\frac{\frac{q^2}{2m}+v_Fq}
{\Omega}\right)^2\right|=\\ \nonumber
&=\frac{m}{\pi q}\ln\left|1+\frac{2v_Fq^3}{m{\Omega}^2}\right|.
\end{align}
Using
\begin{equation}\nonumber
\ln(1+\epsilon)\approx\epsilon\quad\textrm{for}\quad\epsilon\ll1,
\end{equation}
we get
\begin{align}\nonumber
\Pi&\approx\frac{m}{\pi q}\frac{2v_Fq^3}{m{\Omega}^2}=\\
&=\frac{2p_Fq^2}{\pi m{\Omega}^2}.
\end{align}
Let us introduce $D$-dimensional density of states
\begin{equation}
n^{(D)}=2\frac{V_F}{(2\pi)^D}\,,\quad
V_F(p) =\frac{\pi^{D/2}}{(D/2)!}p^D\,,
\quad V_F(p_F)
=2p_F\ \mbox{ for }\ D=1,
\end{equation}
where $V_F$ is the volume of the Fermi sphere. In our case
\begin{equation}
n^{(1D)}=2\frac{2p_F}{2\pi}=\frac{2p_F}{\pi}.
\end{equation}
Then
\begin{equation}
\Pi=\frac{q^2n^{(1D)}}{m{\Omega}^2}.
\end{equation}
Calculating the Fourier transformation, it can be shown
\cite{Stratton,Pendry,Mikhailov} that the photon propagator in long
wavelength approximation ($q\rightarrow0$) obeys
\begin{equation}
V^{(1D)}(\Omega,q)\approx2e^2\ln\left(\frac{1}{qa}\right),
\end{equation}
where $a\geq0\,,\ a\rightarrow0$ is a regularization parameter. More precisely
\begin{equation}
V^{(1D)}(\Omega,q)=e^2\int_{-\infty}^{\infty}\frac{e^{iqx}dx}{|x|}.
\end{equation}
This integral is not convergent -- we have to regularize it
introducing a small diameter of the wire. We give two schemes:
\begin{align}
1)&\quad V^{(1D)}(\Omega,q)=e^2\int_{-\infty}^{\infty}\frac{e^{iqx}dx}{\sqrt{x^2+a^2}}
=-\frac{\pi e^2}{2}(N_0(iaq)+N_0(-iaq))\approx2e^2\ln\left(\frac{2}{\gamma qa}\right),\\
2)&\quad V^{(1D)}(\Omega,q)=2e^2\int_{a}^{\infty}\frac{\cos(x)dx}{x}
=-2e^2\mathrm{Ci}(qa)\approx-2e^2(\gamma+\ln(qa))
=2e^2\ln\left(\frac{1}{e^{\gamma}qa}\right).
\end{align}
Here $\gamma$ is Euler's constant. As it is written above,
both schemes yield
\begin{equation} \nonumber
V\approx2e^2\ln\left(\frac{1}{qa}\right).
\end{equation}
In order to find the dispersion we have to solve the equation
\eqref{eps=0}
\begin{equation}
V\Pi=\frac{2e^2q^2n^{(1D)}}{m{\Omega}^2}\ln\left(\frac{1}{qa}\right)=1.
\end{equation}
As a result, for dispersion of plasma waves we obtain
\begin{equation}
{\Omega}^2(q)=\frac{2e^2 n^{(1D)}}{m}\ln\left(\frac{1}{qa}\right)q^2,
\end{equation}
in agreement with the derivation in \cite{Stratton}.

In the next section we will discuss the case of $D=2$ which is for
first time was discussed in \cite{2DEG}.

\section{2-Dimensional case}
For the case of $D=2$ we introduce polar variables
\begin{align}\nonumber
\mathbf{p}=p(\cos\theta,\sin\theta),\quad\mathbf{q}=q(1,0),\\ \nonumber
d^2p=pdpd\theta,\quad\mathbf{p}\cdot\mathbf{q}=pq\cos\theta,
\end{align}
and
\begin{align}
I_1=2\int\frac{d^2p}{(2\pi)^2}\frac{n_p}
{\frac{p^2}{2m}-\frac{(\mathbf{p}+\mathbf{q})^2}{2m}+\Omega}=
\frac{4m}{(2\pi)^2}\int_0^\infty\int_0^{2\pi}
\frac{p\,\thetaup(p_F-p)dp\,d\theta}{2m\Omega-q^2-2p\,q\cos\theta}.
\end{align}
If we use
\begin{align}
\frac{1}{2\pi}\int_0^{2\pi}\frac{d\theta}{a-b\cos\theta}=
\frac{\mathrm{sign}(a)}{\sqrt{a^2-b^2}},
\end{align}
then
\begin{align}
I_1&=\frac{2m}{\pi}\mathrm{sign}(2m\Omega-q^2)\int_0^{p_F}
\frac{p\,dp}{\sqrt{(2m\Omega-q^2)^2-4p^2q^2}}\nonumber\\
&=\frac{m\,\mathrm{sign}(2m\Omega-q^2)}{2\pi q^2}\left[
|2m\Omega-q^2|-\sqrt{(2m\Omega-q^2)^2-4p_F^2q^2}\,\right].
\end{align}
The velocity on Fermi level is $v_F=p_F/m$, and
\begin{align}
I_1(\Omega,\mathrm q)&=\frac{m^2}{\pi q^2}\left[\;
\Omega-\frac{q^2}{2m}-\mathrm{sign}(\Omega-\frac{q^2}{2m})
\sqrt{\left(\Omega-\frac{q^2}{2m}\right)^2-q^2v_F^2}\,\right],\\
I_2(\Omega,\mathrm q)&=\frac{m^2}{\pi q^2}\left[
-\Omega-\frac{q^2}{2m}+\mathrm{sign}(\Omega+\frac{q^2}{2m})
\sqrt{\left(\Omega+\frac{q^2}{2m}\right)^2-q^2v_F^2}\,\right].
\end{align}
Owing to the symmetry of the polarization operator
$$\Pi(Q)=\Pi(-Q),$$
we obtain for big enough frequency
\begin{align}
\Pi=\frac{m^2}{\pi q^2}\left[-\frac{q^2}{m}+
\sqrt{\left(\Omega+\frac{q^2}{2m}\right)^2-q^2v_F^2}-
\sqrt{\left(\Omega-\frac{q^2}{2m}\right)^2-q^2v_F^2}\,\right].
\end{align}
Working in hydrodynamic approximation \eqref{hydapp}, we introduce the notation
$$\epsilon=\frac{\Omega q^2/m}{\Omega^2+\left(
\frac{q^2}{2m}\right)^2-q^2v_F^2}\ll1.$$
In these terms we have for the polarization operator
\begin{align}
\Pi=\frac{m^2}{\pi q^2}\left[-\frac{q^2}{m}+
\sqrt{\Omega^2+\left(\frac{q^2}{2m}\right)^2-q^2v_F^2}
\left(\sqrt{1+\epsilon}-\sqrt{1-\epsilon}\,\right)\right].
\end{align}
Using
$$\sqrt{1+\epsilon}-\sqrt{1-\epsilon}\approx\epsilon,$$
we arrive at
\begin{align}
\Pi=&\frac{m^2}{\pi q^2}\left[-\frac{q^2}{m}+\frac{\Omega q^2/m}
{\sqrt{\Omega^2+\left(\frac{q^2}{2m}\right)^2-q^2v_F^2}}\right]=\nonumber
\frac{m}{\pi}\left[\frac{\Omega}{\sqrt{\Omega^2-q^2v_F^2}}-1\right]\\
&=\frac{m}{\pi}\left[\frac{1}{\sqrt{1-
\left(\frac{qv_F}{\Omega}\right)^2}}-1\right]=
\frac{m}{\pi}\frac{1}{2}\left(\frac{q\,v_F}{\Omega}\right)^2=
\frac{q^2p_F^2}{2\pi m\Omega^2}.
\end{align}
For the $2D$ case the electron density
\begin{align}
n^{(2D)}=2\frac{\pi p_F^2}{(2\pi)^2}=\frac{p_F^2}{2\pi}.
\end{align}
Then
\begin{align}
\Pi=\frac{q^2n^{(2D)}}{m\Omega^2}.
\end{align}
Again for the dispersion we solve the equation (\ref{eps=0})
\begin{align}
V\Pi=\frac{2\pi e^2qn^{(2D)}}{m\Omega^2}=1.
\end{align}
As a result for dispersion of $2D$ plasma waves, we get
\begin{align}
\Omega^2(q)=\frac{2\pi e^2n^{(2D)}}{m}\,q.
\end{align}

Another important case is the static one $(\Omega=0)$ for $2D$
electron gas where for $q\leq2p_F$ we have
$\mathrm{Re}(\Pi)=-\frac{m}{\pi}$ \cite{2DEG}.

In the next section we will discuss the $D=3$ case which is
described in the monograph by Abrikosov, Gor'kov and Dzyaloshinsky 
\cite{AGD}.

\section{3-Dimensional case}
Analogously for $D=3$ we introduce spherical coordinates
\begin{align}\nonumber
\mathbf{p}=p(\sin\theta\cos\varphi,\sin\theta\sin\varphi,\cos\theta),
\quad\mathbf{q}=q(0,0,1),\\ \nonumber
d^3p=p^2\sin\theta\,dp\,d\theta\,d\varphi,\quad\mathbf{p}\cdot\mathbf{q}=p\,q\cos\theta.
\end{align}
Then (we are not interested in the imaginary part)
\begin{align}
I_1&=2\int\frac{d^3 p}{(2\pi)^3}\frac{n_p}{\frac{p^2}{2m}-
\frac{(\mathbf{p}+\mathbf{q})^2}{2m}+\Omega}\nonumber\\
&= \frac{4m}{(2\pi)^3}\int_0^\infty\int_0^\pi\int_0^{2\pi}
\frac{p^2\,\thetaup(p_F-p)\sin\theta\,dp\,d\theta\,d\varphi}
{2m\Omega-q^2-2p\,q\cos\theta}\\
&=\frac{m}{2\pi^2q}\int_0^{p_F}p\ln\left|
\frac{2m\Omega-q^2+2p\,q}{2m\Omega-q^2-2p\,q}\right|dp\nonumber.
\end{align}
To simplify the notations we introduce
$a=2m\Omega-q^2,\ b=2q,$ so
\begin{align}
\int_0^{p_F}p\ln\left|\frac{a+bp}{a-bp}\right|dp&=
\frac{1}{2}p_F^2\ln\left|\frac{a+bp_F}{a-bp_F}\right|-
ab\int_0^{p_F}\frac{p^2dp}{a^2-b^2p^2}\nonumber\\
&=\left(\frac{1}{2}p_F^2-\frac{a^2}{2b^2}\right)
\ln\left|\frac{a+bp_F}{a-bp_F}\right|-\frac{a}{b}p_F,
\end{align}
and
\begin{align}
I_1&=\frac{m}{4\pi^2q}\left(p_F^2-\frac{(2m\Omega-q^2)^2}{4q^2}\right)
\ln\left|\frac{2m\Omega-q^2+2p_F\,q}{2m\Omega-q^2-2p_F\,q}\right|\\
&\quad+(2m\Omega-q^2)\frac{mp_F}{4\pi^2q^2},\nonumber\\
I_2&=-\frac{m}{4\pi^2q}\left(p_F^2-\frac{(2m\Omega+q^2)^2}{4q^2}\right)
\ln\left|\frac{2m\Omega+q^2+2p_F\,q}{2m\Omega+q^2-2p_F\,q}\right|\\
&\quad-(2m\Omega+q^2)\frac{mp_F}{4\pi^2q^2}.\nonumber
\end{align}
Finally we obtain
\begin{align}\label{Pi3D}
\Pi&=\frac{m}{4\pi^2q}\left(p_F^2-\frac{(2m\Omega-q^2)^2}{4q^2}\right)
\ln\left|\frac{2m\Omega-q^2+2p_F\,q}{2m\Omega-q^2-2p_F\,q}\right|\\
&-\frac{m}{4\pi^2q}\left(p_F^2-\frac{(2m\Omega+q^2)^2}{4q^2}\right)
\ln\left|\frac{2m\Omega+q^2+2p_F\,q}{2m\Omega+q^2-2p_F\,q}\right|-
\frac{mp_F}{2\pi^2}.\nonumber
\end{align}
The Fourier transformation of the static polarization operator gives the well
known from the physics of magnetism RKKY interaction. Technical details are given
in Ref.~\onlinecite{RKKY}.

Introducing the notations
\begin{align}
g(\Omega)&=\frac{m(\Omega^2-q^2v_F^2)}{2q^3v_F}
\ln\frac{\Omega+qv_F}{\Omega-qv_F},\qquad\Omega_\pm=\Omega\pm q^2/2m,\\
\Omega_e&=\sqrt{4\pi n^{(3D)}e^2/m},\qquad n^{(3D)}=2\frac{
\frac{4}{3}\pi p_F^3}{(2\pi)^3}=\frac{p_F^3}{3\pi^2},
\end{align}
the longitudinal polarizability takes the form \cite{Lifshitz10}
\begin{equation}
\varepsilon_l(\Omega,q)-1=\frac{3\Omega_e^2}{2q^2v_F^2}
(1-g(\Omega_+)+g(\Omega_-)).
\end{equation}

As we did before, we use the hydrodynamical approximation \eqref{hydapp}, and
\begin{equation}
\ln\left|\frac{1+x}{1-x}\right|\xrightarrow[x\to0]{}2x+\frac{2}{3}x^3,
\qquad x=\frac{2p_F\,q}{2m\Omega\pm q^2}.
\end{equation}
That is why equation \eqref{Pi3D} transforms into
\begin{align}
\Pi&=\frac{m}{4\pi^2q}\left(p_F^2-\frac{(2m\Omega-q^2)^2}{4q^2}\right)
\left(\frac{4p_F\,q}{2m\Omega-q^2}+
\frac{16p_F^3q^3}{3(2m\Omega-q^2)^3}\right)\nonumber\\
&-\frac{m}{4\pi^2q}\left(p_F^2-\frac{(2m\Omega+q^2)^2}{4q^2}\right)
\left(\frac{4p_F\,q}{2m\Omega+q^2}+
\frac{16p_F^3q^3}{3(2m\Omega+q^2)^3}\right)-
\frac{mp_F}{2\pi^2}\nonumber\\
&=\left(1-\frac{1}{3}\right)\frac{mp_F^3}{\pi^2}
\left(\frac{1}{2m\Omega-q^2}-\frac{1}{2m\Omega+q^2}\right)\\
&+\frac{4mp_F^3q^2}{3\pi^2}\left(\frac{1}{(2m\Omega-q^2)^3}-
\frac{1}{(2m\Omega+q^2)^3}\right)
=\frac{p_F^3q^2}{3\pi^2m\Omega^2}.\nonumber
\end{align}
Taking into account the density of states we can write the polarization operator in
another form
\begin{align}
\Pi=\frac{q^2n^{(3D)}}{m\Omega^2}.
\end{align}
This hydrodynamic asymptotic is valid for arbitrary dimensions.

Turning back in \eqref{eps=0}
\begin{align}
V\Pi=\frac{4\pi e^2}{q^2}\frac{q^2n^{(3D)}}{m\Omega^2}=1,
\end{align}
we finally obtain the well-known result for bulk plasma
oscillations
\begin{align}
\Omega^2(q)=\frac{4\pi e^2n^{(3D)}}{m}=\mathrm{const}.
\label{jk6sb}
\end{align}

\acknowledgments
One of the authors (TM) is thankful to  Tz. Sariisky for
introducing him in the quantum field methods \cite{Ceco} in solid state plasma
physics.

\end{document}